\documentclass[numberedappendix,12pt,preprint]{emulateapj}

\usepackage{amsmath,amssymb,amsxtra,amsfonts}
\usepackage{graphicx}
\usepackage{color}
\definecolor{midgray}{gray}{0.4}
\definecolor{orange}{rgb}{1,0.5,0}
\usepackage[colorlinks=true,citecolor=midgray,linkcolor=midgray]{hyperref} 
\usepackage{bm}

\shorttitle{Multi-wavelength Analysis of Dust Emission}
\shortauthors{Yue N. et al.}

\begin{document}

\title{A new method to quantify differentiate collapse models of star formation}

\author{Nannan Yue\altaffilmark{1,2*}, Di Li\altaffilmark{1,3} and Zhiyuan Ren\altaffilmark{1}}
\affil{
$^1$ National Astronomical Observatories, CAS, Beijing 100012; \\
$^2$ University of Chinese Academy of Sciences, Beijing 100049, China\\
$^3$ Key Laboratory of Radio Astronomy, Chinese Academy of Science, Nanjing, 210008, China
}

\email{{*}Email: yuenannan0907@gmail.com}

\begin{abstract}
Continuum emissions from dust grains are used as a general probe to constrain the initial physical conditions of molecular dense cores where new stars may born.
To get as much information as possible from dust emissions, we have developed a tool, named as $COREGA$, which is capable of identifying positions of dense cores, optimizing a three-dimensional model for the dense cores with well characterized uncertainties. 
$COREGA$ can also estimate the physical properties of dense cores, such as density, temperature, and dust emissivity, through analyzing multi-wavelength dust continuum data sets.
In the numerical tests on $COREGA$, the results of fitting simulated data are consistent with initial built-in parameters.
We also demonstrate $COREGA$ by adding random gaussian noises with Monte Carlo methods and show that the results are stable against varying observational noise intensities within certain levels. 
A beam size $<$ 3 arcsec and rms $<$ 0.2mJy/pixel (1 pixel = 0.1") is needed for ALMA to distinguish different collapse models, such as power law and Bonner-Ebert sphere, during continuum observations of massive dense cores in Orion molecular cloud.
Based on its advanced algorithm, $COREGA$ is capable of giving a quick and deep analysis on dust cores.

\end{abstract}


\section{Introduction}
Star formation is always a key step to understand our universe, from the initial beginning of simple particles to the beautiful evolution of complex structures.
As the potential star forming site, dense molecular cores which are mainly embedded in cold dark clouds, are significant objects to study. 
The physical structures of them are important to the following born stars and up to galaxies. 
Dust, as an important probe of cores' structures, has been well studied in the last decades.
Multi-wavelength submillimeter and millimeter images of dust emissions from dense cores have been obtained from both ground based telescopes and space instruments. 
With PACS and SPIRE instruments, several Herschel surveys have obtained multi-wavelength continuum data, which cover the peak of Spectral Energy Distributions (SEDs) from dense cores. 
Besides, large wavelength coverage and good spatial resolution of ALMA make it to have the potential in providing unprecedented insights into the density profile, temperature structure, and dust properties of dense cores.
These insights from analyzing ALMA data will give a much clearer picture of the initial conditions of star formation, the dynamical states of cores, and the mass distribution of cores as well as its relevance to the stellar initial mass function (IMF).

Density profile is a crucial probe of the energy state and the key dynamic process (e.g. collapse) in star formation. 
It quantifies core's structure through comparison with theories of core evolution in different stages.

For a self-gravitating, isothermal sphere within which internal pressure everywhere precisely balances the inward push of gravity and external surface pressure, a Bonner-Ebert sphere is naturally used to describe density profile.
It only has a single dependable parameter $\xi_{max}$, a dimensionless, characteristic radius.
 Bonnor (1956) and Ebert (1955) investigated when $\xi_{max}>6.5$ such pressure truncated clouds will gravitational collapse. 
Quantities of observations on starless and protostellar cores (Alves et al. 2001; Lada et al., 2004; Teixeira et al., 2005; Kandori et al., 2005; Myers 2005) and numerical simulations are typically found to exhibit Bonner-Ebert density profiles.

These profiles have a systematic manner, evolving from a relatively flat, equilibrium structure to a highly condensed collapsing structure.
Besides the hydrostatic equilibrium solution, classical self-similar solution is provided by Shu (1977). 
Within this model, the mass distribution throughout the core becomes more centrally concentrated as the prestellar cores evolve.
 The gravitational ``inside out" collapse of isothermal spheres at constant accretion rate, happens, which start as gas clouds not far removed from the condition if marginal stability. 
 An $r^{-2}$ law holds for the density distribution in the static outer envelope and an $r^{-3/2}$ law for the freely falling inner envelope. 
 These power law profiles seem mathematically simple but are never observationally confirmed, because of the limited telescope resolution to distinguish a flat region from singular ones. 
In addition to the above isothermal, gravitational spheres, the evolution becomes quite different in the non-isothermal core (Foster \& Chevalier 1993), or in situations involving nonthermal pressure support such as turbulence or magnetic fields (see, for example, Ward-Thompson, Motte, \& Andr'e 1999). 
All these models make different predictions about the form of the radial density profiles in a dense core. 
High resolution and sensitivity are needed to determine density profiles to distinguish varies models.
That's why we need ALMA.

The temperature structure of cores provides critical information regarding the heating source. A particular interesting case would be a cold core with a hot central region, but without infrared source. It is very desirable to have effective ways of searching for such sources, which are presumably heated by gravitational contraction and thus at the very onset of forming new stars. That is the so-called first hydrostatic cores.
For the parameter of dust emissivity, it can be an important sign of core evolution. There have been evidences of emissivity variation in star formation regions, which may be related to grain growth. We can use multi-band data sets to constrain properly the fitting of dust emissivity from SEDs and give a more accurate estimation of dust emissivity spectral index distribution (Schnee et al. 2014).

To differentiate these star formation models, we developed a self-consistent program COREFGA.
Through analyzing multi-band dust continuum data, we can obtain the physical properties of dense cores, and obtain a three dimensional model for the dust cores with well characterized uncertainties. 
$COREGA$ can generate simulated images based on assumed density and temperature distribution through dust radiative transfer and thus provides an opportunity for a quantified comparison between assumed density-profile models and observed multi-band images.
It also gives the capability requirements for telescopes to distinguish different density profiles predicted by different collapse models.

In this paper, we focus on the algorithm and principles of $COREGA$ in Sec. {\ref{sec:aa}} . 
Then we will give the numerical tests on the forward generator, an important component of the tool in Sec. {\ref{sec:tests}}. 
In Sec. {\ref{sec:noise}} and Sec. {\ref{sec:alma}}, the analysis of noise effect and continuum predictions for ALMA observation are shown separately.  
Summary and discussion are present in Sec. {\ref{sec:con}}.

\section{COREGA}
\begin{figure}
\includegraphics[width=0.5\textwidth]{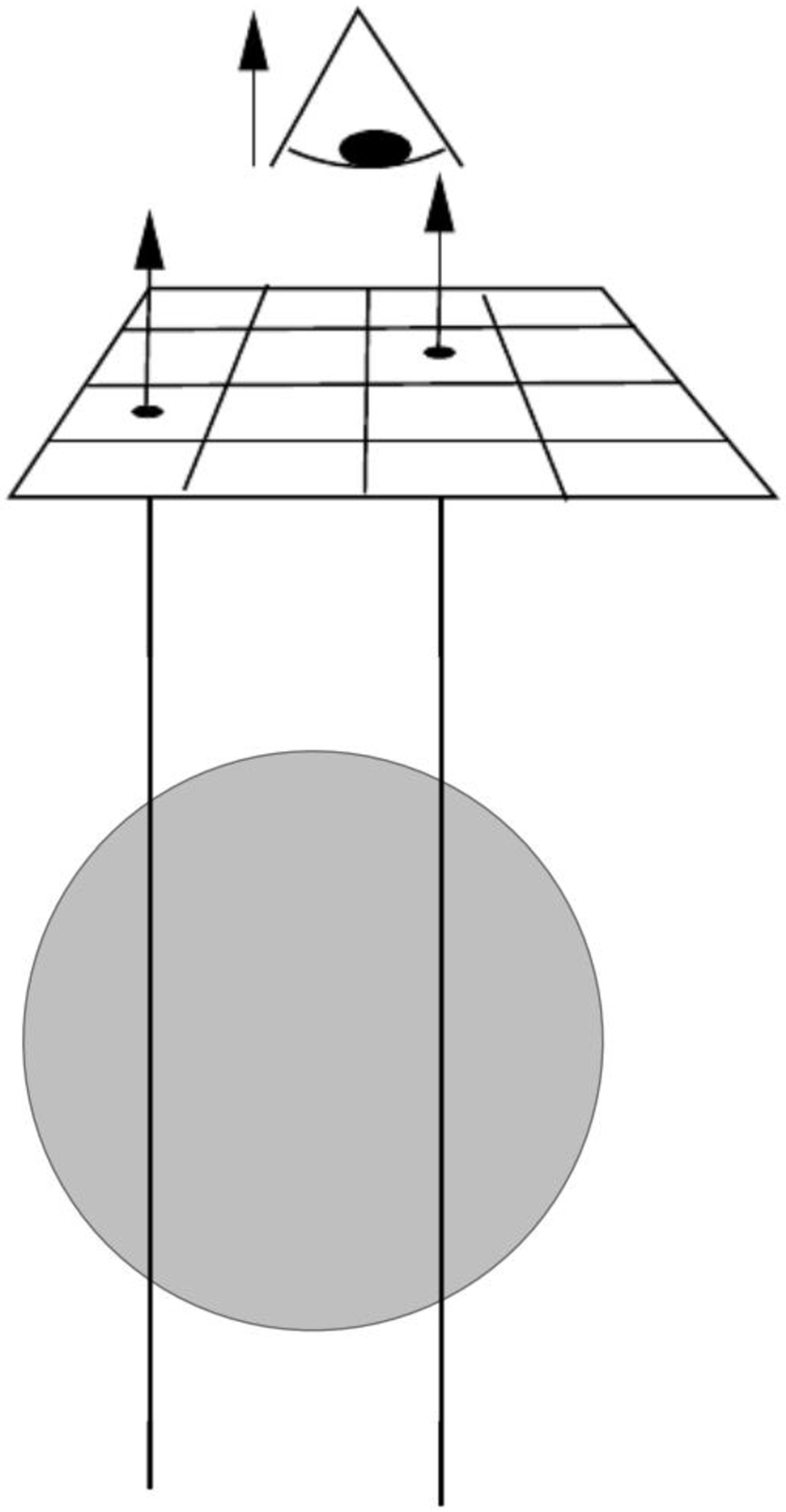}
\caption{The schematic diagram showing the photons transfer through a molecular core to an observer in the right side.}
\label{fig:show}
\end{figure}

{\label{sec:aa}}
COREGA is a tool applicable to three-dimensional core modeling.
The main purpose of $COREGA$ is to estimate profiles of density and temperature of a cold core using observed images at a set of wavelengths and spectral energy distributions of locations. 
The schematic diagram is shown in Figure {\ref{fig:show}}.
The tool has three components.

1. \textit{The forward generator}, which generates simulated images based on the given density and temperature profiles. The simulator assumes an onion-like shell structure and performs a radiative transfer calculation to obtain the output image.

2. \textit{A core finding tool}, which extracts the positions of cores based on multi-wavelength information.

3. \textit{The iterative profiling tool}, outlined above, which utilizes the forward simulator to derive the 3-d structure of cores.

\subsection{Data Preprocess}
Multi-wavelength images of dust continuum emission, the input of the procedure, may be obtained by different telescopes with various resolutions and sensitivities. Hence, data preprocess, which includes regridding the images, is needed. 
To make optimal use of the available emission intensity, we can utilize higher spatial resolution at shorter wavelengths to compensate the poorer resolution at longer wavelengths, since the algorithm deals with data at multiple wavelengths simultaneously. Some super resolution (a factor of 2-3) is provided by the use of these prior information.
 
We can give the position of core as a input value and can also use the procedure to find the core. Firstly, find the position with maximal flux as the location of core's center. Then, do the fine adjustment by shifting the image based on a least squares fit with a reference Point Spread Function (PSF).

\subsection{Model}
\begin{figure}
\includegraphics[width=0.5\textwidth]{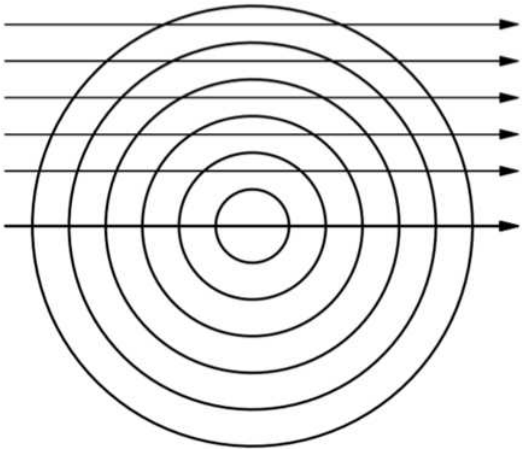}
\caption{The schematic diagram of photon's path in the core.}
\label{fig:rad}
\end{figure}

{\label{sec:model}}
We build up a radially symmetric 3-dimensional core model in our procedure $COREGA$. It's an onion-like shell structure as shown in Figure {\ref{fig:rad}}. Temperature and density profiles depend on layer radius. They are described by the following equations.
\begin{eqnarray}
\label{eq:t}
T(r)=T_{1}+\frac{T_{0}-T_{1}}{1+(\frac{r}{r_t})^2},
\end{eqnarray}

\begin{eqnarray}
\label{eq:n}
n_{H_{2}}(r)=\frac{n_{H_{2}}(0)}{1+(\frac{r}{r_0})^\alpha},
\end{eqnarray}

where $T_{0}$, $T_{1}$, $r_0$, $r_t$, $\alpha$, and $n_{H_{2}}(0)$ are quantities we use to parametrize the radial profiles of temperature and density. The examples  of density and temperature are shown in Figure {\ref{fig:n}} and Figure {\ref{fig:t}}, representatively.

Detailed calculation of radiation transfer is shown in Appendix A. Along with another parameter $\beta$, the emissivity spectral index, used in the equation (\ref{eq:q}), these seven parameters are necessary for fitting in $COREGA$. We can also estimate the core mass $M$, through the distribution of molecular hydrogen density $n_{H_{2}}$.

The default volume density model used in the procedure is based on the work of \citet{2004A&A...416..191T} . 
Appendix A of \citet{2004A&A...416..191T} illustrated the rationality of this analytical expression of equation (\ref{eq:n}). 
This form is widely used in establishing density model.
In the equation (\ref{eq:n}), $n_{H_{2}}(0)$ is the central density, $r_0$ is the radius of the inner "flat" region, and 
$\alpha$ is the asymptotic power index.

Theoretical models of the dust temperature in cores predict a slight inward decrease (e.g., \citet{2001ApJ...557..193E} ). We choose a temperature profile able to describe this prediction, shown as the equation (\ref{eq:t}). It can describe two thermal cases through changing responding parameters. The first case is a core with a cold central region, hotter outside. The second case is a core with a hot central region and decreasing temperature outside. The second one would be a particular interesting case, if the core has a hot central region but without inside infrared source. It may be the candidate of the first cores. Examples are shown in Figure \ref{fig:n} and \ref{fig:t}.

\begin{figure}
\includegraphics[width=0.5\textwidth]{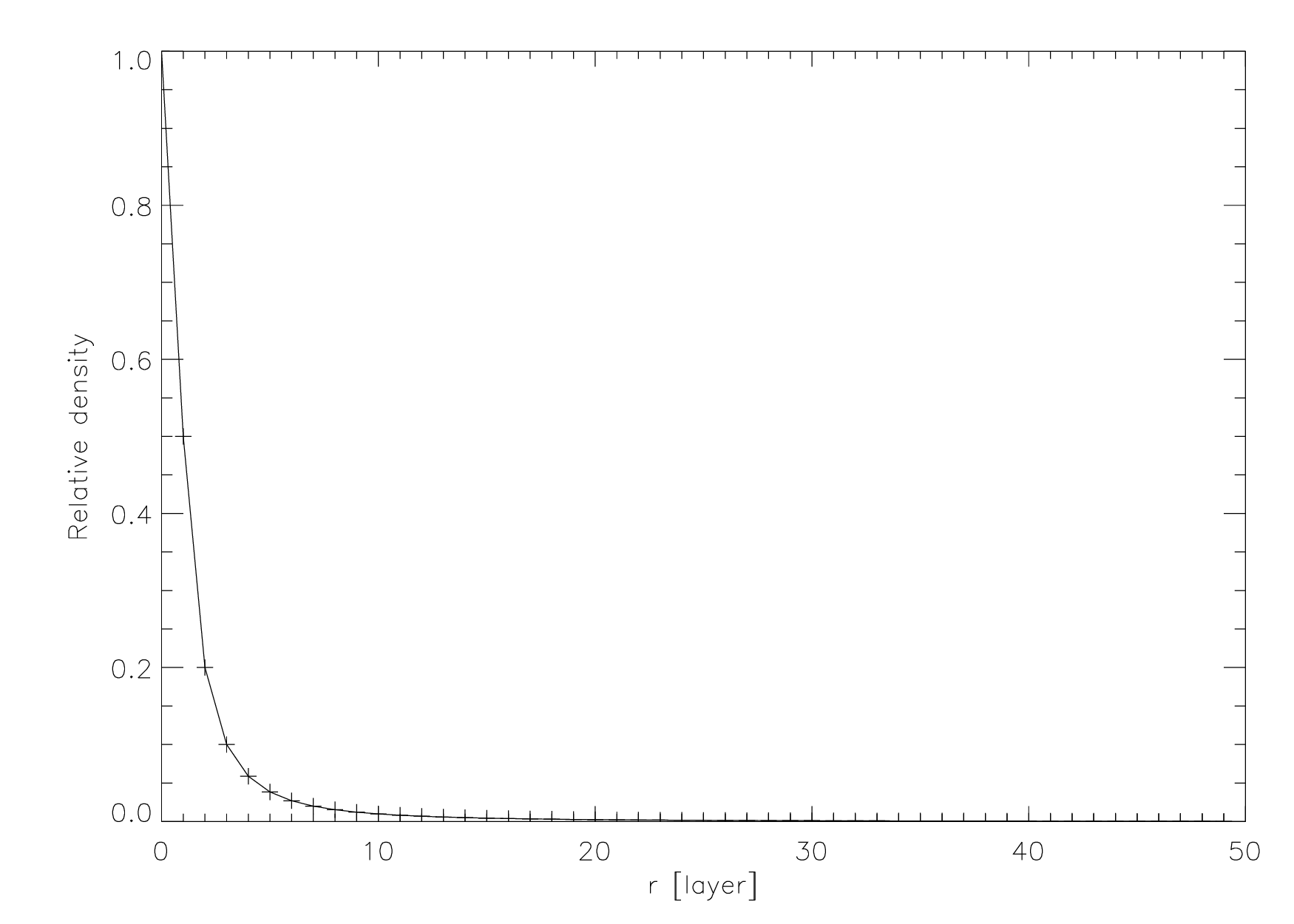}
\caption{An example of relative density profile. $n_{H_{2}}(r)/n_{H_{2}}(0)$ with $r_0=5$ and $\alpha=1.6$.  }
\label{fig:n}
\end{figure}

\begin{figure}
\includegraphics[width=0.5\textwidth]{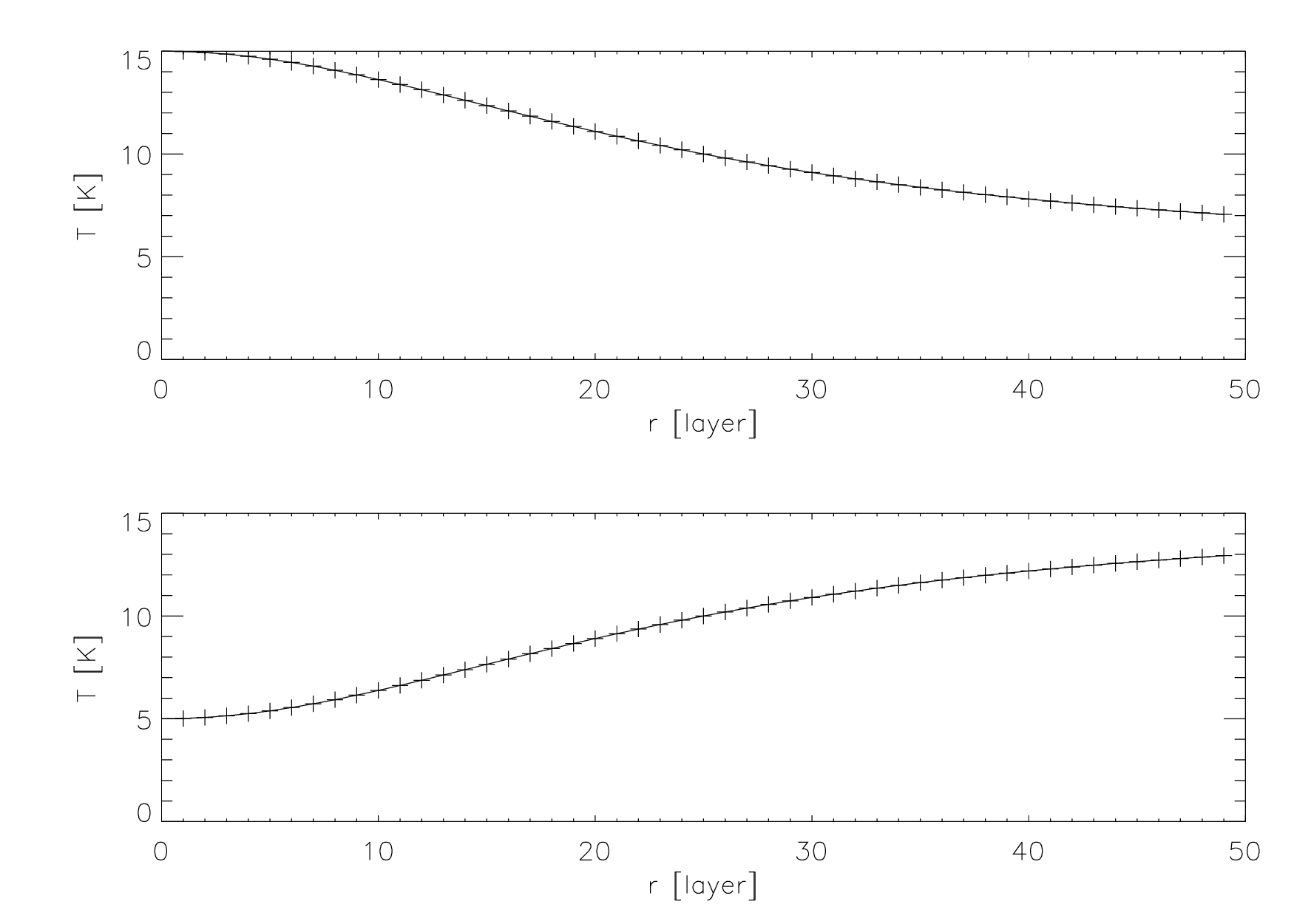}
\caption{An example of temperature profile. The $upper$ panel shows a core with a colder central region (the first case), at the values $T_{0}=15$, $T_{1}=5$, $r_t=25$. The $lower$ panel shows a core with a hotter central region (the second case), at the values $T_{0}=5$, $T_{1}=15$, $r_t=25$.}
\label{fig:t}
\end{figure}

The procedure is enough flexible. The temperature and density profiles can be changed easily by modifying it in the procedure, leading to different models with different profiles. Thus $COREGA$ can be applied to distinguish different core models, by analyzing the responding $\chi ^2$ between models.

%
%

\subsection{Algorithm and Principles}
COREGA is based on Bayesian algorithm. It combines multi wavelength continuum data with its corresponding PSF to extract temperature and density structure in the cores.
The algorithm obtains the most probable solution for the radial distributions of dust temperature and density. 

We considered two different measurement models, appropriate to two different regimes of the measurement system.
In the first case, which involves non-parametrized radial profiles, we solve for the temperature and density at  each radial distance independently. This is appropriate for well-solved cores.
In the second case, appropriate to less well-solved cores, we parametrize the radial profiles of temperature and density. We can also incorporate other parameters of interests, such as spectral index of the opacity law at one or more radial positions in the core. 
In either case, our approach makes optimal use of the available spatial information. 
We choose to parametrize the radial profiles of temperature and density to deal with even less well-solved cores in $COREGA$. 
Given that the cores are barely resolved at longer wavelengths, we believe that our approach extracts the maximum possible information from the observed data. 
Supplementary data with high angular resolution at lower wavelengths will enhance greatly the core extraction.

Our analysis involve fitting a radially-symmetric core model, with parametrized temperature and density.  The distributions of dust temperature, and density are assumed to be described by 1-dimensional functions of the radial distance from the star, as shown in equations (\ref{eq:t}) and (\ref{eq:n}). We then obtained a maximum likelihood solution for the set of radial samples of temperature and density by maximizing the conditional probability given by equation (\ref{eq:p}):

\begin{eqnarray}
\label{eq:p}
\ln P(\bm{z}|\bm{b}) = -\frac{1}{2}\sum_{k,m}[b_{km}-(\bm{H}_{k}*\bm{M}_{k}(\bm{z})_{m})]^2/\sigma^{2}_{km}\nonumber
\\+const
\end{eqnarray}

where $\bm{z}$ is a vector whose components are the unknowns ($T_{i}$, $n_{i}$, $\beta$, e.t ) and $\bm{b}$ is the measurement vector whose components consist of the pixel values of the observed images at all wavelengths. $H_{k}$ represents the PSF at the $k^{th}$wavelength.
$M_{k}(z)$ represents the theoretical intensity distribution projected onto the plane of the sky for particular set of model parameters, and $*$ denotes convolution. As shown in Figure \ref{fig:rad}, we do the radiative transfer calculation assuming the onion-like structure with parameterized temperature \ref{eq:t} and density \ref{eq:n}.
$\sigma_{km}$ represents the measurement noise in the $m^{th}$ pixel at the $k^{th}$ wavelength, 

The solution is obtained by numerical maximization of the above equation, subject to positivity constraints on the values of temperature and optical depth. 
It is based on the Newton-Raphson technique using initial parameters obtained by running a grid of models. The initial estimate of parameters is got by coarse minimization and then, do the finial minimization of $-\ln P$ using Powell procedure, which is already involved in IDL lib. 
Powell's method is an algorithm for finding a local minimum of a function by a bi-directional search along each search vector, in turn. The method is useful for calculate the minimum of a continuous but complex function, especially one without  an underlying mathematical definition, because it is not necessary to take derivatives\citep{powell1964}.

%
%


\section{Forward Generator Test }{\label{sec:tests}}
To make sure the procedure reasonable physically to analyze cold cores, we do several simple forward generator numerical tests through changing different physical parameters in the models. 

\subsection{Effect of mass and wavelength}
\begin{figure}
\includegraphics[width=0.5\textwidth]{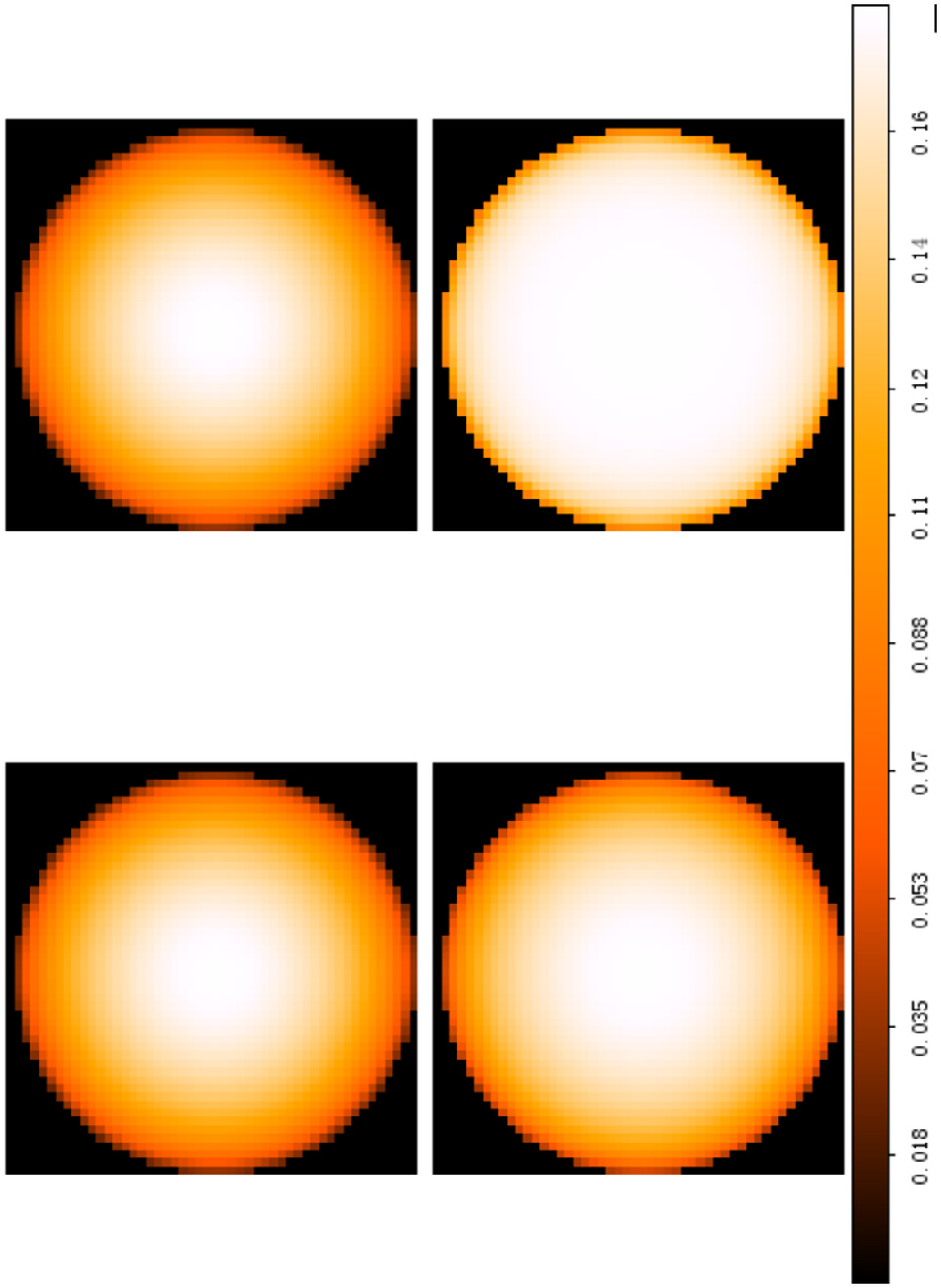}
\caption{Simulated dense core images of different masses, (clockwise from top left) 1, 10, 240, and 1000 $M_{\odot}$. The images are of cores with uniform temperature and density profiles at the wavelength of $100\mu m$. The color is shown in linear between min and max value in different images, not absolute. The color bar is only for the case of 1000 $M_{\odot}$.}
\label{fig:dm}
\end{figure}

Firstly, consider the simplest case, in which cores have uniform temperature ($T=15K$) and relative density (${n_{H_{2}}(r)}/{n_{H_{2}}(0)}=1$). If cores have different masses, their absolute density is different and then optical depths, which reflect in flux images at certain wavelength directly. Hence, we firstly test core mass as a variable in producing flux images in our procedure. The spatial intensity distributions of different cores are shown in Figure {\ref{fig:dm}}.
Assume the core has a radius of 0.1 $pc$ at the distance of 140 $pc$ from us, just like the core TMC-1C \citep{2005ApJ...624..254S}. The  emissivity spectral index $\beta$ is taken as the usual value of $-2$. 
The masses are chose as 1, 10, 240, and 1000 $M_{\odot}$ separately.The opacity depth through the center of a core with a mass 240 $M_{\odot}$ at $100\mu m$ is around $1$.
As shown in Figure {\ref{fig:dm}}, when opacity depth is bigger enough (such as the case of 1000 $M_{\odot}$), the flux distribution is more homogeneous. That is for a large optical depth, which trends to infinite, the equation (\ref{eq:irte}) in LTE approaches the Planck blackbody radiation only dependent on temperature.

\begin{figure}
\includegraphics[width=0.5\textwidth]{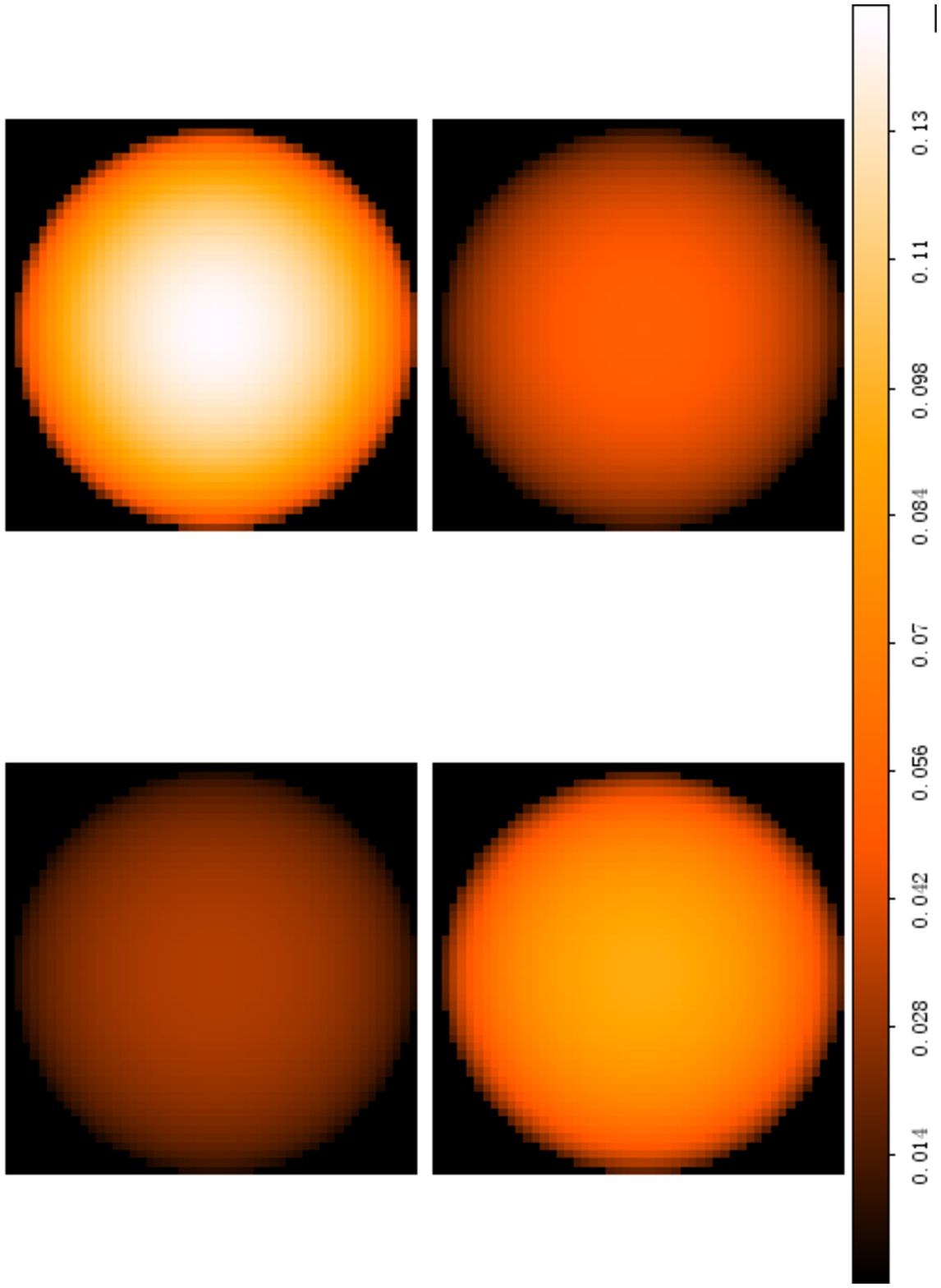}
\caption{Simulated dense core images at different wavelengths, (clockwise from top left) 100, 200, 300, and 400 micron.
The modeled core is a 50 $M_{sun}$ core with a uniform density profile and temperature of 15K.
}
\label{fig:lamb}
\end{figure}

\begin{figure}
\includegraphics[width=0.5\textwidth]{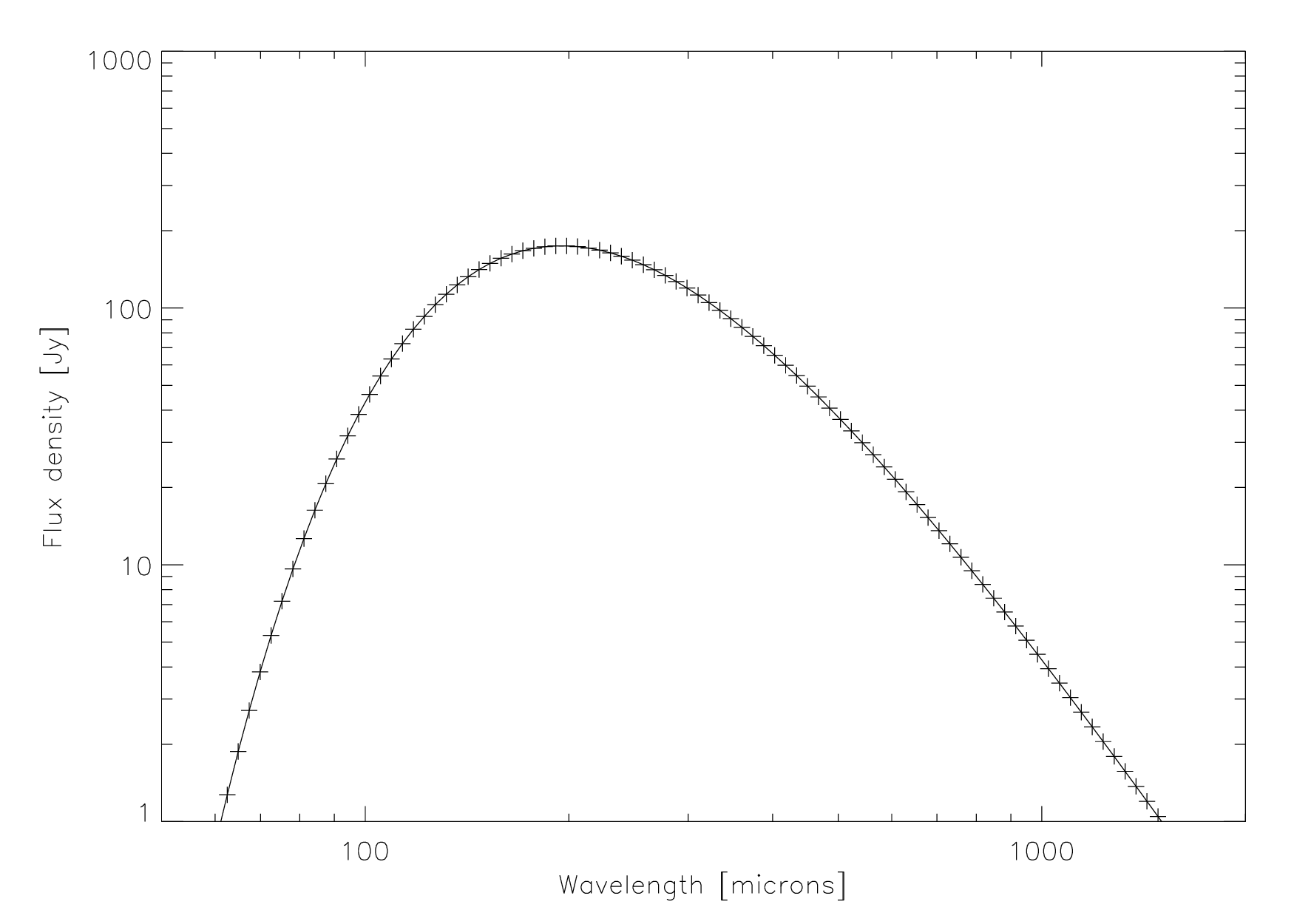}
\caption{The simulated sed of dense core. The modeled core has a mass of 50 $M_{sun}$ and the density and temperature profiles are as the same as Figure {\ref{fig:lamb}}.}
\label{fig:sed}
\end{figure}

The second test is to produce models of a core's radiative transfer at different wavelengths. 
As shown in Figure \ref{fig:lamb}, $200\mu m$ is brighter than others. For a black body in thermodynamic equilibrium, there is \textit{Wien's displacement law} in which 15K responses to $200\mu m$ for the maximal flux. The total simulated spectral energy distribution of the whole core is shown in Figure \ref{fig:sed}. We also give the SED of different locations shown in Figure. 

%
%

\subsection{Temperature}
\begin{figure}
\includegraphics[width=0.5\textwidth]{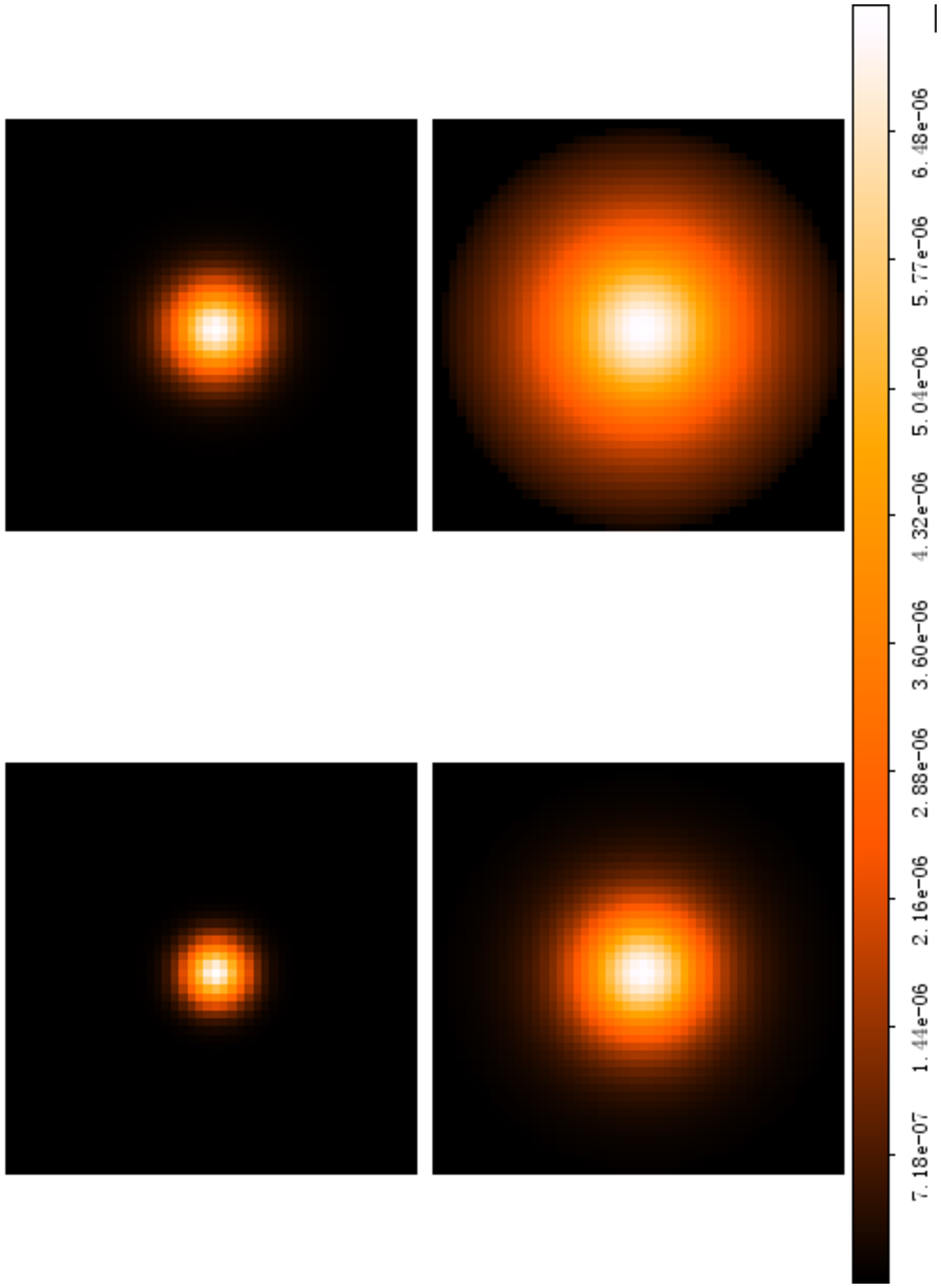}
\caption{Simulated dense core images at (clockwise from top left) 50, 100, 200, and 400 $\mu m$.
The modeled core is an externally heated 50 $M_{sun}$ core with a uniform density profile and temperature as the upper part of Figure {\ref{fig:t}}. The color is shown in linear between min and max value in different images, not absolute. The color bar is only for the case of 50 $\mu m$.}
\label{fig:t1}
\end{figure}

\begin{figure}
\includegraphics[width=0.5\textwidth]{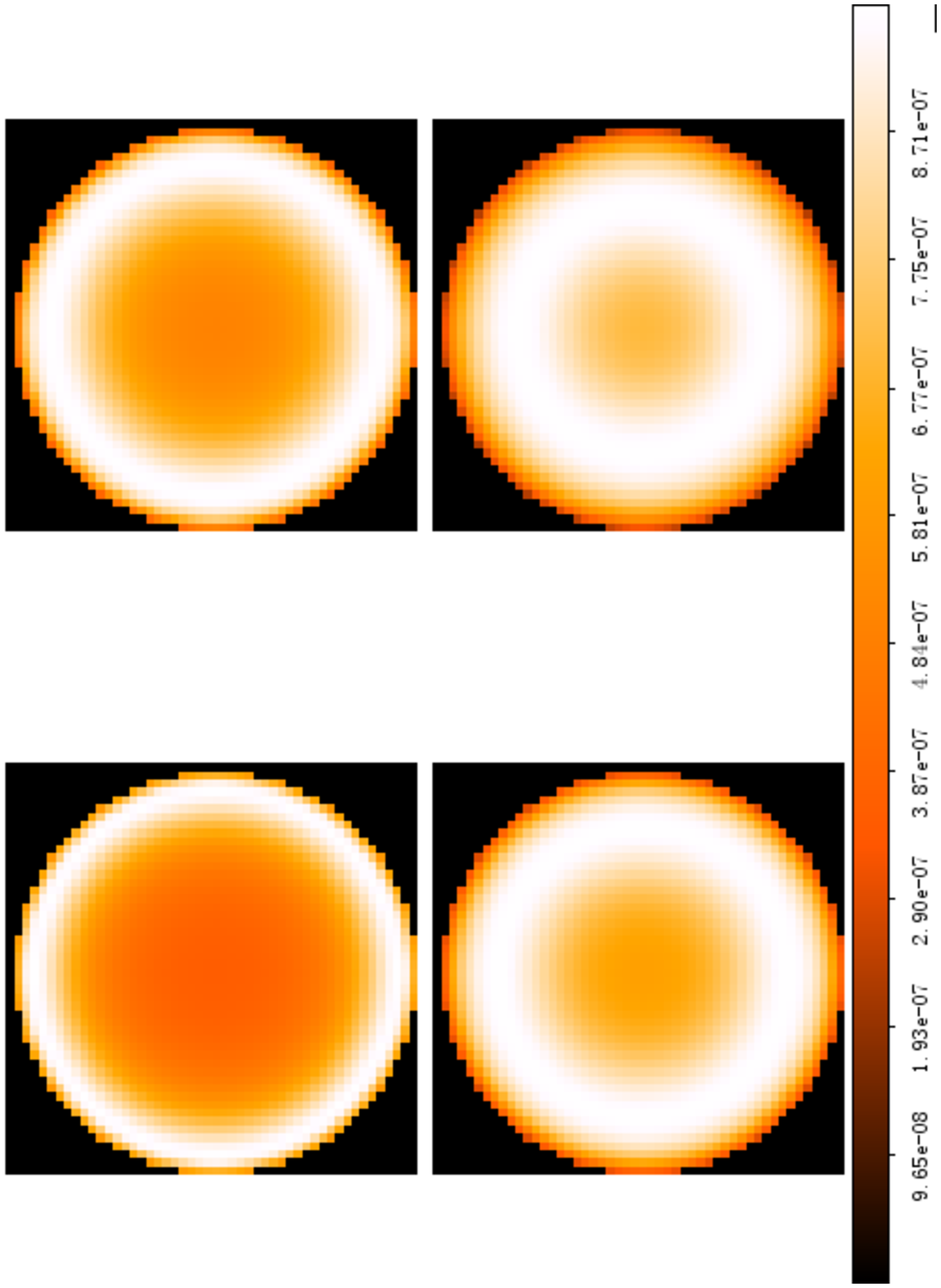}
\caption{Simulated dense core images at (clockwise from top left) 50, 100, 200, and 400 $\mu m$.
The modeled core is an externally heated 50 $M_{sun}$ core with a uniform density profile and temperature as the lower part of Figure {\ref{fig:t}}. The color is shown in linear between min and max value in different images, not absolute. The color bar is only for the case of 50 $\mu m$.}
\label{fig:t2}
\end{figure}

For different thermal structures, we can model seperately to understand more directly.
Using the temperature in Figure \ref{fig:t}, the simulated dense core images are shown in Figure \ref{fig:t1} and \ref{fig:t2}. When the temperature decreases from the center, the core trends darker outside shown in Figure \ref{fig:t1} and when the temperature increases, limb-brightened effects show in Figure \ref{fig:t2}.

%
%

\subsection{Dependence of Core Emission on model parameters}{\label{sec:par}}

As introduced in Sec. {\ref{sec:model}}, there are seven parameters in all, $T_{0}$, $T_{1}$, $r_0$, $r_t$, $\alpha$, $n_{H_{2}}(0)$, and $\beta$. COREGA's aim is to find the optimal solutions of these seven parameters. 
In order to learn the dependence among these parameters, we give simple tests on a single parameter while fixing others. 
Figures are given below, in which y axis represents ${\chi}^2$ quantifying the goodness of solution. The red points in these figures are the true values built in the model.

We can find the shapes of $\alpha$, $mass$, and $\beta$ are similar, which are all have relatively systemically decrease towards the true values.  $mass$ and $\beta$ are more sensitive to the ${\chi}^2$, while $\alpha$ changed less slowly.

For $T_{0}$ , $T_{1}$, $r_0$, the values larger than the true value have sharply increased ${\chi}$, which means the solutions solved is less impossible larger than the instinctive true values. The small figures embedded in the large ones are zooming in around the true value. It shows $T_{0}$ is less constrained than $T_{1}$, which has a sharp turn around the true value in the zooming-in figure. The $r_t$ profiles shows the opposite trend, which decreases more sharply in the smaller edge.

\begin{figure}
\includegraphics[width=0.5\textwidth]{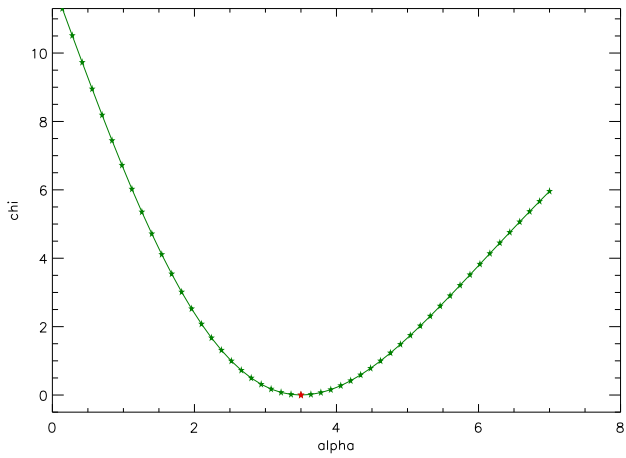}
\caption{$\alpha$ vs  ${\chi}^2$ }
\label{fig:alpha}
\end{figure}

\begin{figure}
\includegraphics[width=0.5\textwidth]{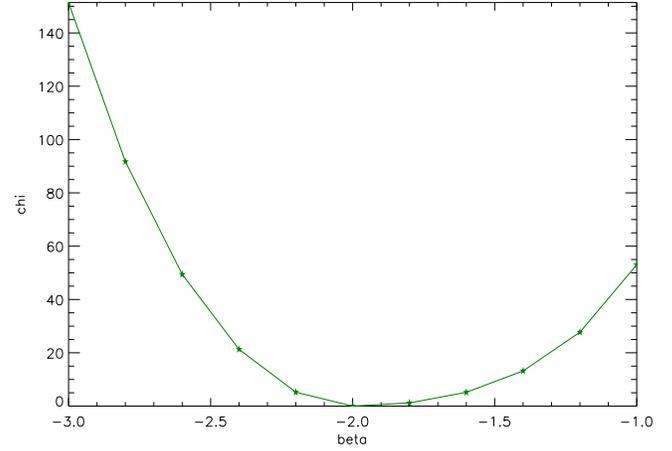}
\caption{$\beta$ vs  ${\chi}^2$ }
\label{fig:beta}
\end{figure}

\begin{figure}
\includegraphics[width=0.5\textwidth]{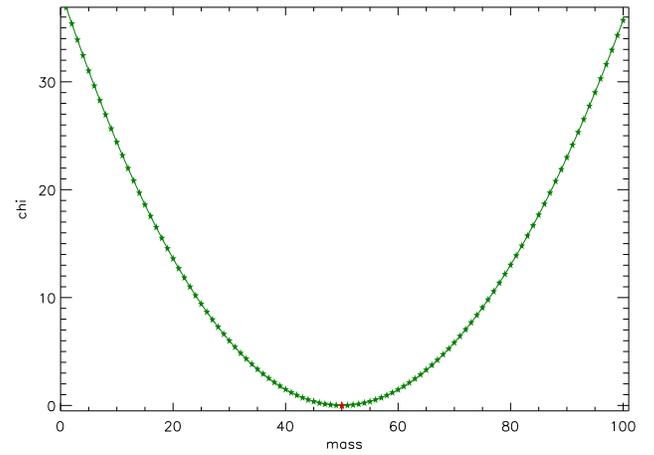}
\caption{$mass$ vs  ${\chi}^2$ }
\label{fig:mass}
\end{figure}

\begin{figure}
\includegraphics[width=0.5\textwidth]{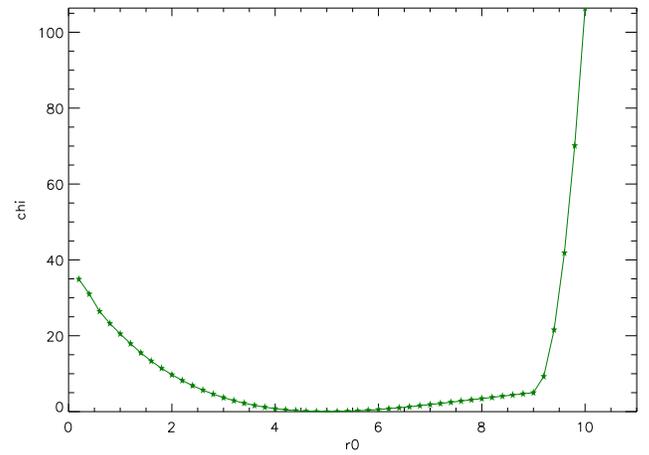}
\caption{$r_0$ vs  ${\chi}^2$ }
\label{fig:r0}
\end{figure}

\begin{figure}
\includegraphics[width=0.5\textwidth]{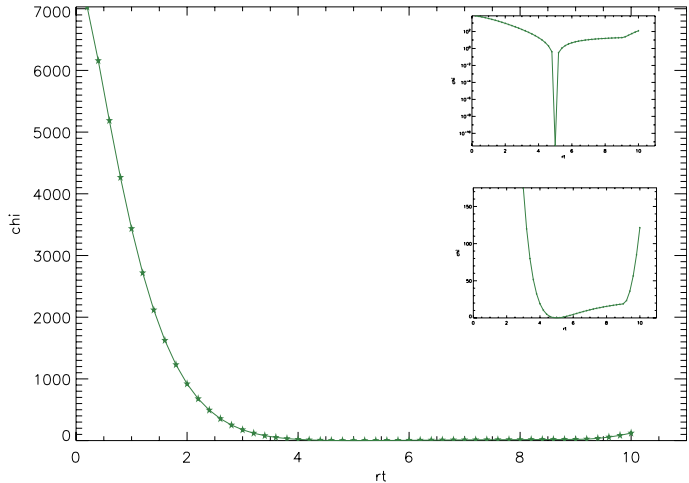}
\caption{$r_t$ vs  ${\chi}^2$ }
\label{fig:rt}
\end{figure}

\begin{figure}
\includegraphics[width=0.5\textwidth]{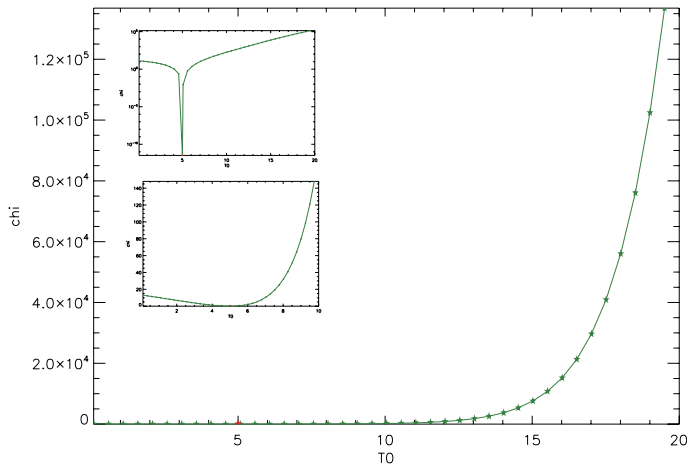}
\caption{$T_0$ vs  ${\chi}^2$ }
\label{fig:T0}
\end{figure}

\begin{figure}
\includegraphics[width=0.5\textwidth]{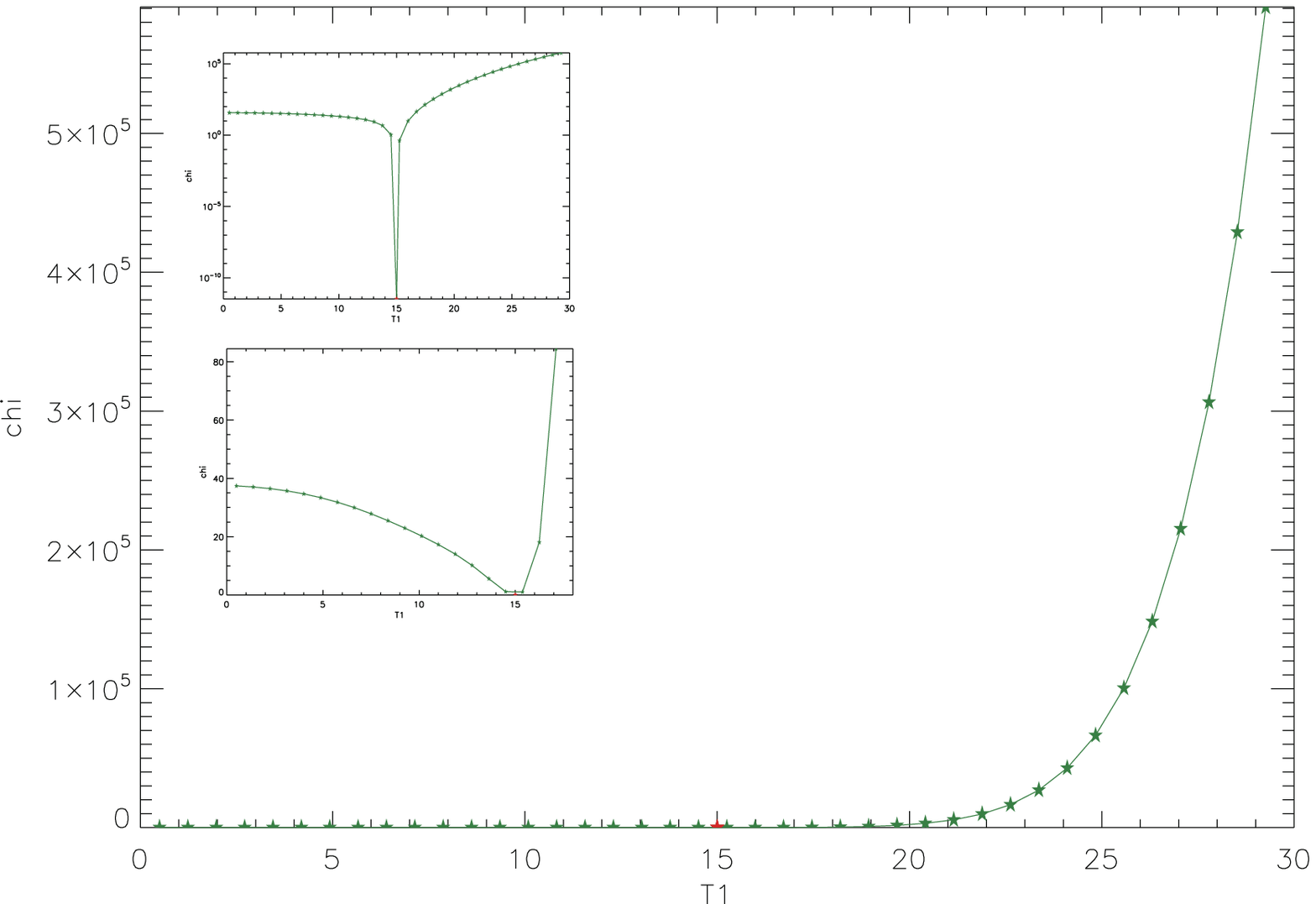}
\caption{$T_1$ vs  ${\chi}^2$ }
\label{fig:T1}
\end{figure}

%
%

\section{Effect of Random Measurement Noise}{\label{sec:noise}}

\begin{figure}
\includegraphics[width=0.5\textwidth]{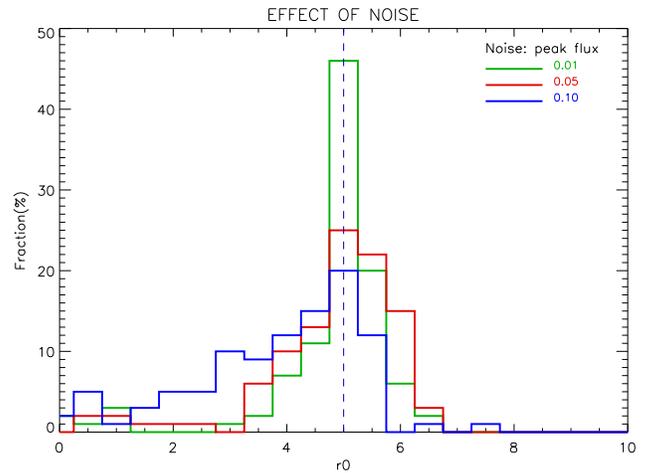}
\caption{Monte Carlo test of $COREGA$ robustness to increasing noise. Different line colors represent different RMS noise values in the fraction of peak flux. }
\label{fig:no}
\end{figure}

To demonstrate the robust of the procedure against random noises, we test it using a Monte Carlo simulation. We add different levels of Gaussian noises into model images and use the procedure to fit the profiles.
 The fitting success of seven parameters are defined as the fraction of successfully-fitted values out of 100 trials. Here shows one of temperature parameter $T_1$ distribution in Fig {\ref{fig:no}}. It shows that $COREGA$ is stable within certain level, such as noise minimizes factor 5 of peak flux.

%
%




%
%

\section{producing synthetic ALMA observations}
\begin{figure}
\includegraphics[width=0.5\textwidth]{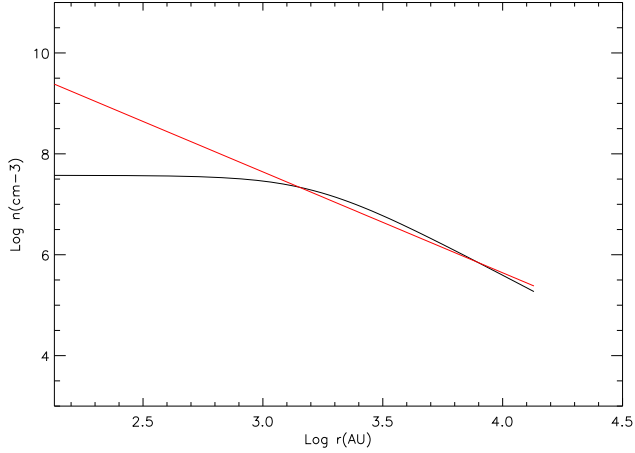}
\caption{The two types of density profile used for the simulated observation. The red line is power law profile and the black one Bonner Ebert Profile.}
\label{fig:dbp}
\end{figure}

{\label{sec:alma}}
As discussed above, density profiles within the cores provide important constraints in distinguishing between star formation models. 
We frame the essential observational question for the massive cores, which could be in supercritical collapse rather than hydrostatic equilibrium, in terms of differentiating between a BE profile and a power-law profile.
To predict different forms of the radial density profiles, the key aspect of effective observational tests is to measure the density profile in a spatial range close to the core center. 
There have to be multiple beams with efficient S/N to differential BE and power-law type density profile shown in Fig {\ref{fig:dbp}}..
We will discuss these two observational measurements' impact on resolving density models.

\begin{figure}
\includegraphics[width=0.5\textwidth]{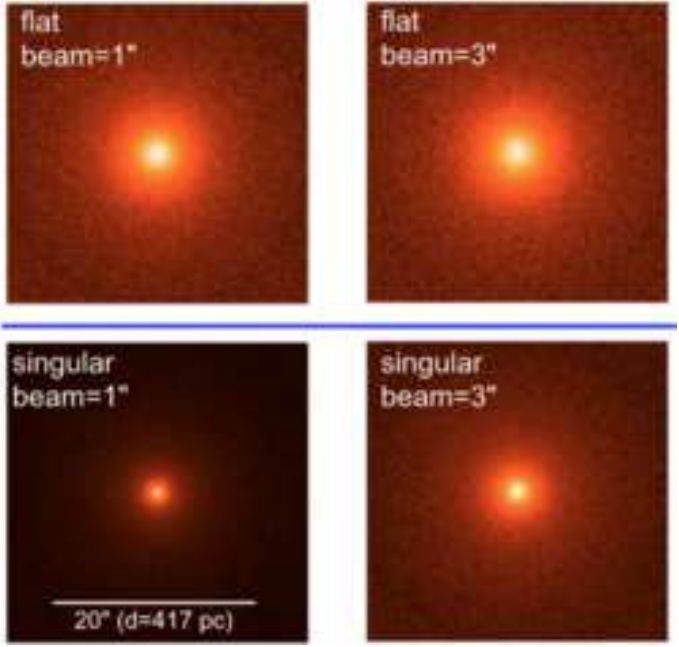}
\caption{The simulated 1.0 mm continuum observation of a dust core with 40 $M_{\odot}$ at the same distance with Orion (D = 417 pc). Two different density profiles and at two angular resolutions are simulated using COREGA program with constant temperature T = 20 K.}
\label{fig:ib}
\end{figure}

\begin{figure}
\includegraphics[width=0.5\textwidth]{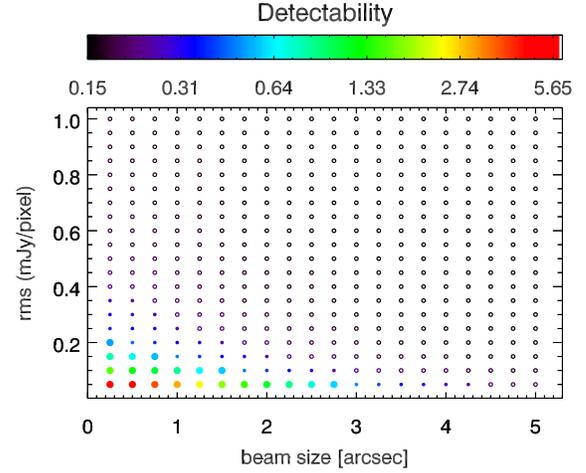}
\caption{The probability that a power-law density profile (central-peaked) can be significantly distinguished from a flat profile in COREGA. 
A larger $\chi$  value suggests that two types of profiles can be more significantly distinguished. A reasonable recipe for ALMA is suggested to be rms ~0.1 mJy/beam and beam 1.000". And once the beam gets larger, the power to resolve the density profiles is quickly getting weak.}
\label{fig:fn}
\end{figure}

A tailor-made test case is designed for ALMA observation of a massive molecular core with a mass of $40 M_{sun}$ at the distance of Orion molecular cloud.
The fluxes at $1000 \mu m$  convolved with different beam sizes are shown in Fig{\ref{fig:ib}}, which demonstrates that the ability to distinguish power-law from BE of a massive core depends sensitively on resolution. The Fig{\ref{fig:fn}} gives more quantities analysis of model detectability based on different resolutions and noises. The defined detectability are calculated by the equation (\ref{eq:p}):
\begin{equation}
\label{eq:chi}
{\chi ^2} = \sum\limits_k {\frac{{{{({x_{c,k}} - {y_{c,k}})}^2}}}{{{\sigma^2}(N - 1){{(1 + \frac{r}{{{r_0}}})}^2}}}}
\end{equation}

A beam size $<$ 3 arcsec and rms $<$ 0.2mJy/pixel (1 pixel = 0.1") is needed to distinguish this certain core.

\section{Applications on real observations}
\begin{figure}
\includegraphics[width=0.5\textwidth]{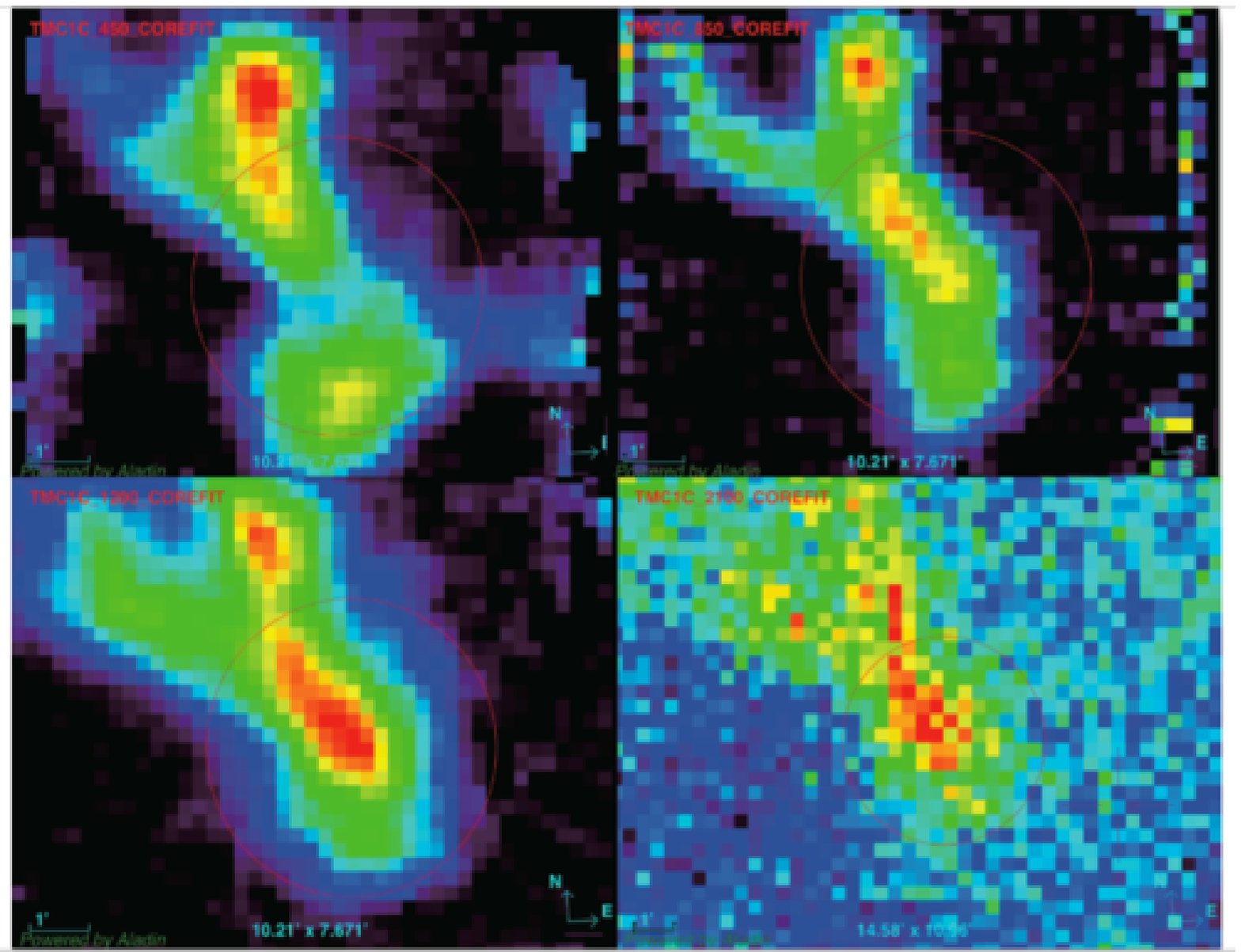}
\caption{The 450, 850 and 1200 and 2100 $\nu$m emission maps of TMC-1C. The red circle shows the location of the core.}
\label{fig:tmcct}
\end{figure}

\begin{figure}
\includegraphics[width=0.5\textwidth]{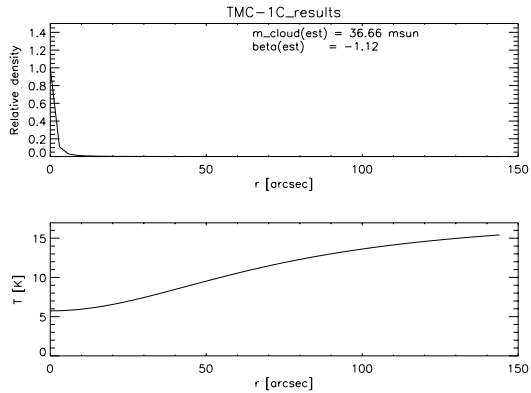}
\caption{The temperature and density structures of TMC-1C resolved by COREGA.}
\label{fig:tmcrst}
\end{figure}

{\label{sec:realcase}}
We take TMC-1C as an example to test COREGA. TMC-1C is a starless core in the Taurus molecular cloud at an approximate distance of 140pc. Previous studies ( Schnee et al. 2005, Schnee et al. 2010) have determined that it is cold and dense at its center, and becomes less dense and warmer at larger radii. The temperature and density profiles we obtain is shown in Fig{\ref{fig:tmcrst}}. The dust emission continuums used to do the fitting are shown in Fig{\ref{fig:tmcct}}. Our work is consist with the ones down by Schnee et al. in 2005 \& 2010.

\begin{figure}
\includegraphics[width=0.5\textwidth]{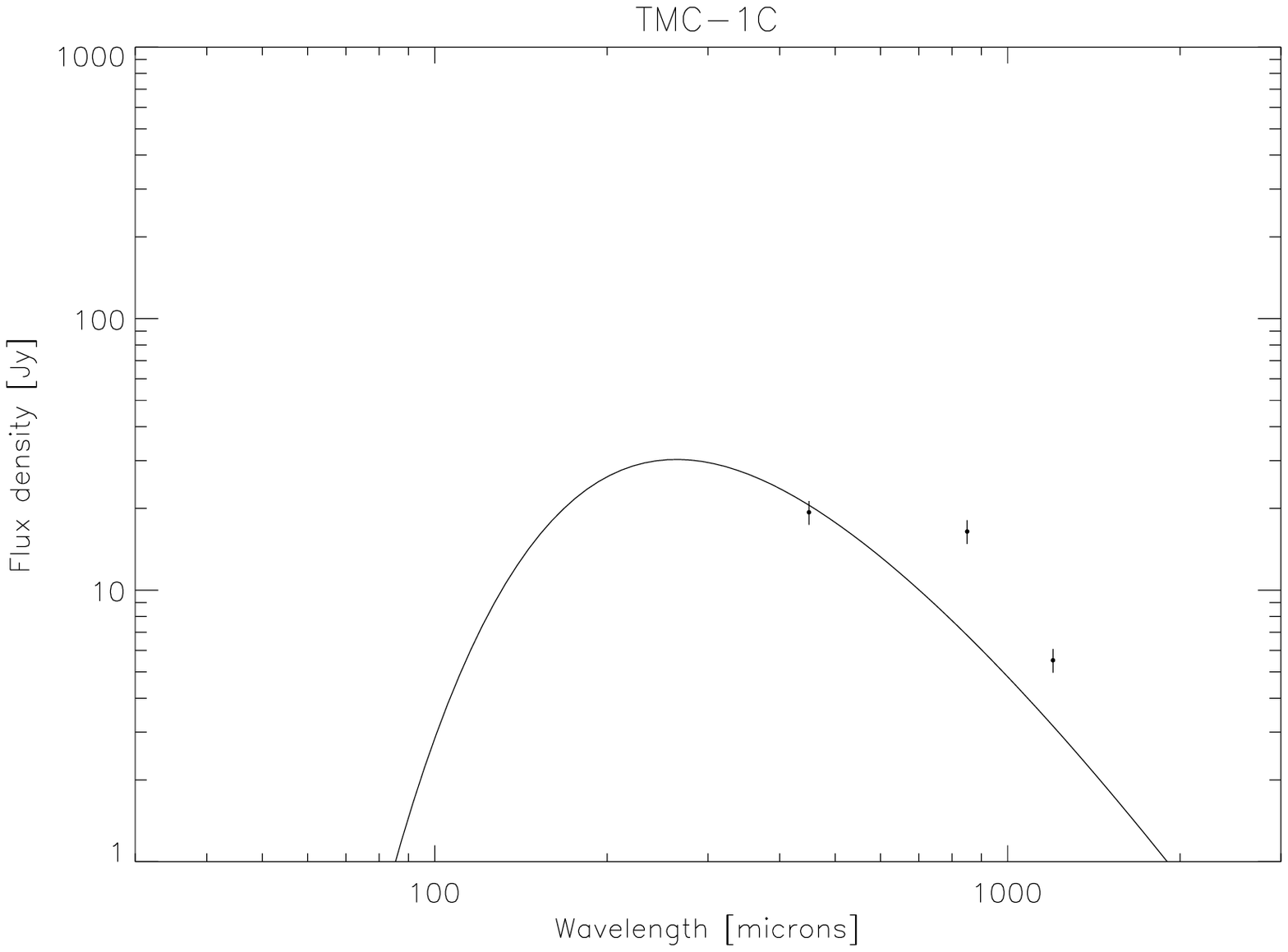}
\caption{The simulated sed. The cross symbol shows the real data including the sed data and continuum data.}
\label{fig:tmcsed}
\end{figure}

\begin{figure}
\includegraphics[width=0.5\textwidth]{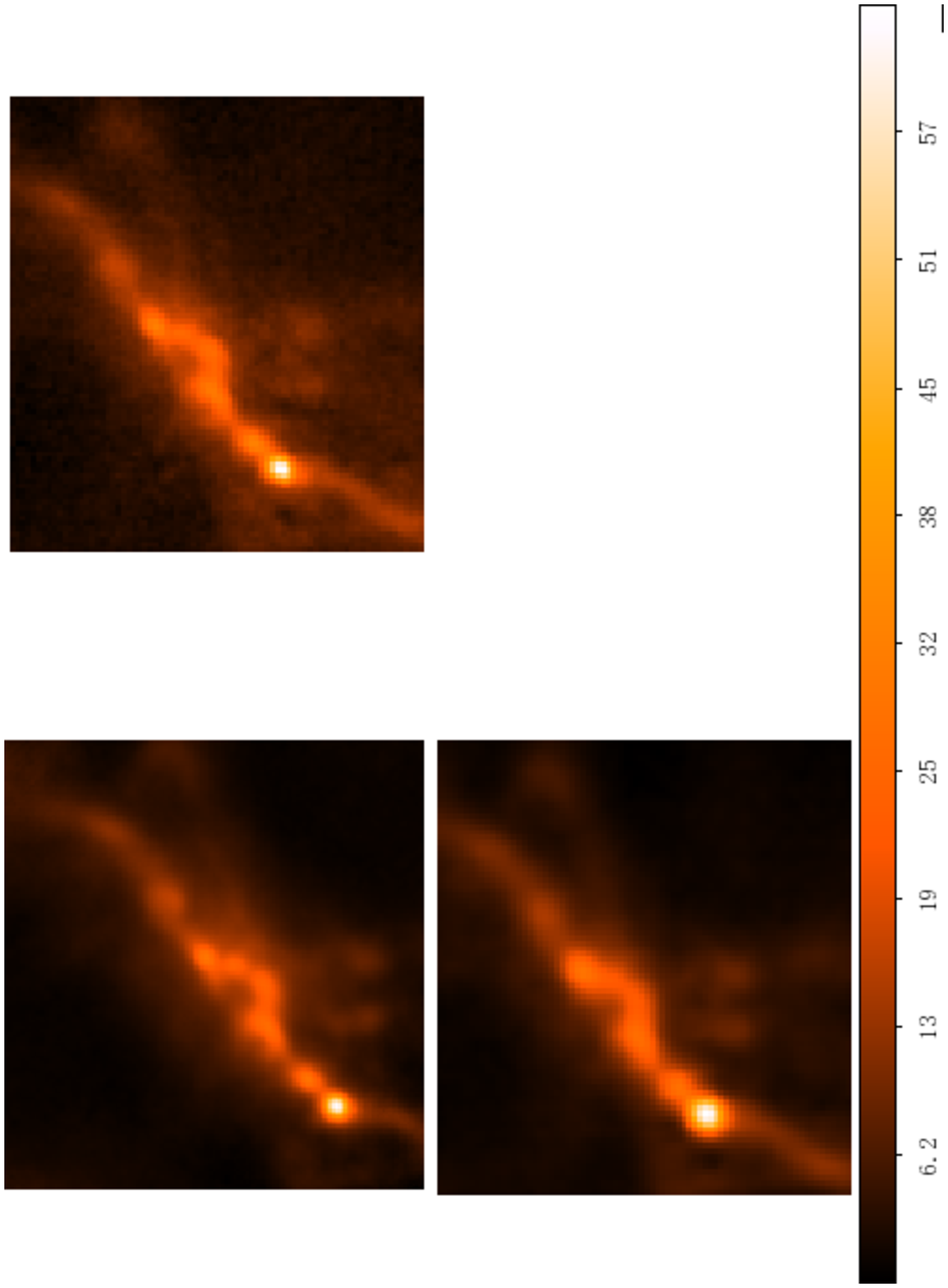}
\caption{The 350, 450 and 850?m emission maps of part of Orion3.}
\label{fig:ori1-13full}
\end{figure}

\begin{figure}
\includegraphics[width=0.5\textwidth]{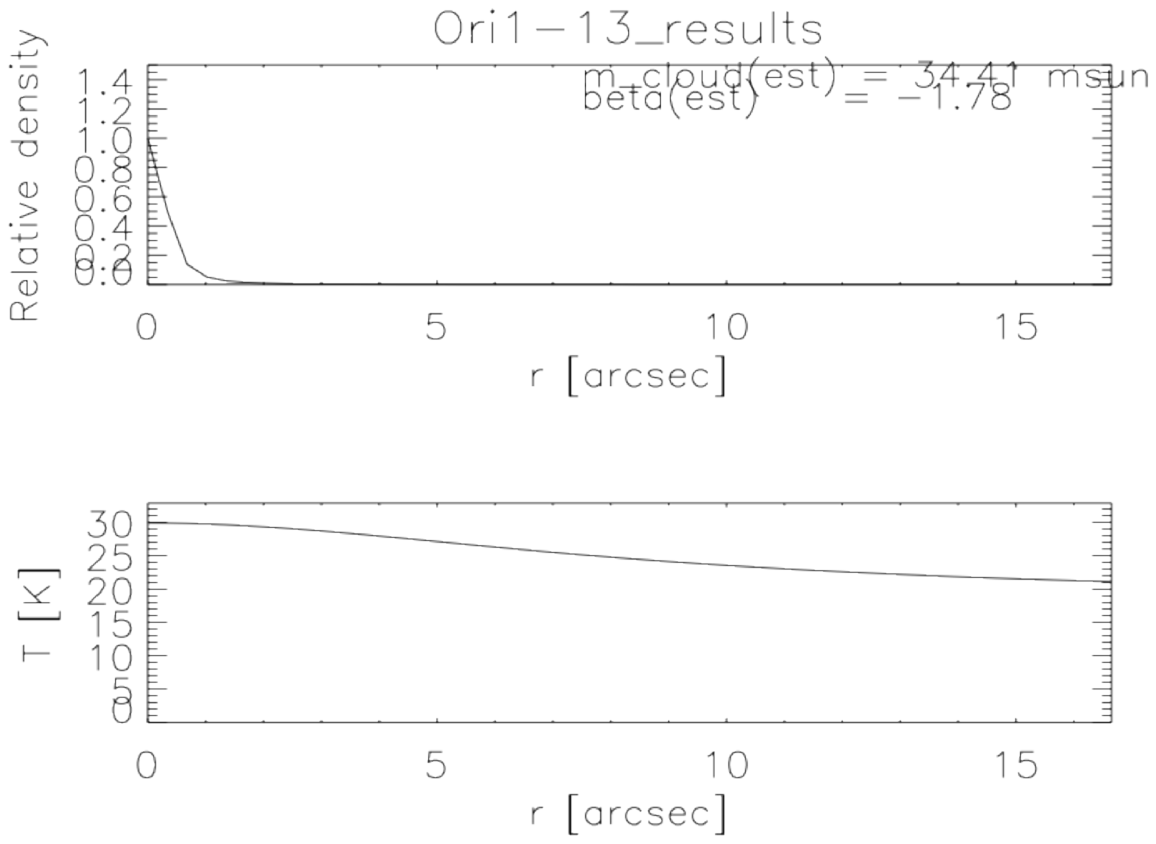}
\caption{The temperature and density structures of ORI1-13 resolved by COREGA.}
\label{fig:ori1-13}
\end{figure}

We take another core in massive star formation of Orion Molecular Cloud, named as MMS 6.  
The temperature and density profiles we obtain is shown in Fig{\ref{fig:ori1-13}}. The dust emission continuums used to do the fitting are shown in Fig{\ref{fig:ori1-13full}}. Our work is consist with the ones down by Takahashi et al. in 2007. 
The bolometric luminosity, temperature, and core mass of MMS 6, which were derived from the previous single-dish millimeter to submillimeter observations combined with IRAS data, are $<$60 L$_{\odot}$, 15-25 K, and 36 M, respectively (Chini et al. 1997). 
This brightest source is located at the center of the OMC-3 region, and the 1.3 mm flux is roughly one order of magnitude larger than those for any other continuum sources in OMC-2/3 (Chini et al. 1997; Johnstone \& Bally 1999). Despite the unusual appearance of MMS 6, no signature of star formation activities such as molecular outflow or jet has been detected toward MMS 6 by Takahashi et al. (2008a).

\begin{figure}
\includegraphics[width=0.5\textwidth]{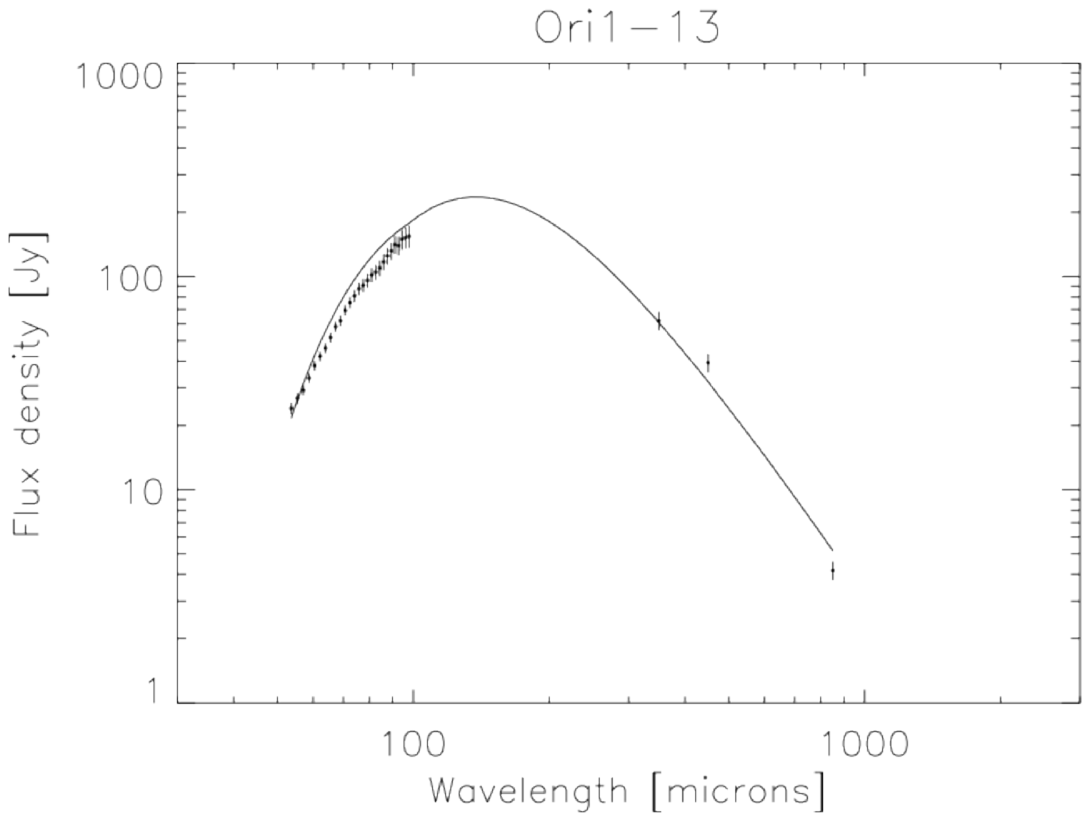}
\caption{The simulated sed. The cross symbol shows the real data including the sed data and continuum data.}
\label{fig:ori1-13sed}
\end{figure}

\section{Summary and Conclusions}{\label{sec:con}}
COREGA is an effective method to analyze the dust emissions in dense cores, which shows very stable numerical behavior. The three dimension structure of the cores, is the first time to be obtained, by resolving the integrated line of sight emission from the core using accurate simulations of different band images. Sensitivity and resolution are key conditions to reveal the inner part of density profiles, which is important for further core collapse models.
For massive dense cores in orion molecular cloud, a beam size $<$ 3 arcsec and rms $<$ 0.2mJy/pixel (1 pixel = 0.1") is needed to detect between Bonner Ebert profile and power law profile.
Based on advanced telescope technique, such as ALMA, we hope to distinguish different collapse mechanisms and improve the star formation theories in the future.


\acknowledgments
This work was supported by the Ministry of Science and Technology National 
Basic Science program (project 973) under grant No. 2012CB821800.

\appendix

\section{Radiative Transfer Calculation}
To obtain the number of photons reaching to us, radiative transfer must be build in our procedure.
Considering a cluster of photons pass through matter, energy may be added or subtracted by emission or absorption, and the specific intensity $I$ will not remain constant in general. 
The variation of specific intensity can be described by the following transfer equation:
\begin{eqnarray}
\frac{dI_{\nu}}{d\tau_{\nu}}=-I_{\nu}+S_{\nu},
\end{eqnarray}	
where the $\tau_{\nu}$ is the optical depth, and $S_{\nu}$ stands for the source function defined as the ratio of emission coefficient to absorption coefficient. By regarding all quantities as functions of optical depth, we can formally solve the radiative transfer equation.
Multiply the equation by the integrating factor $exp(-\tau_{\nu})$, we can get the formal solution of the transfer equation:
\begin{eqnarray}
\label{eq:s}
I_{\nu}(\tau_{\nu})=I_{\nu}(0)e^{-\tau_{\nu}}+\int_{0}^{\tau_{\nu}}e^{-(\tau_{\nu}-\tau'_{\nu})}S_{\nu}(\tau'_{\nu})d\tau'_{\nu}.
\end{eqnarray}
The above equation can be regarded as the sum of two terms: the initial intensity diminished by absorption plus the integrated source.
Assuming a constant source function $S_{\nu}$ not dependent on $\tau$ , equation (\ref{eq:s}) gives the solution:
\begin{eqnarray}
\label{eq:ss}
I_{\nu}(\tau_{\nu})=I_{\nu}(0)e^{-\tau_{\nu}}+S_{\nu}(1-e^{-\tau_{\nu}}).
\end{eqnarray}

In the case of thermodynamic equilibrium, Kirchhoff's law for thermal emission is used.
\begin{eqnarray}
S_{\nu}=B_{\nu}(T).
\end{eqnarray}	
Then equation (\ref{eq:ss}) becomes:
\begin{eqnarray}
\label{eq:rte}
I_{\nu}(\tau_{\nu})=I_{\nu}(0)e^{-\tau_{\nu}}+B_{\nu}(T)(1-e^{-\tau_{\nu}}).
\end{eqnarray}
Here $B_{\nu}$ is the Planck function.
\begin{eqnarray}
\label{eq:p}
B_{\nu}(T)=\frac{2h\nu^3}{c^2}\frac{1}{e^{h\nu/kT}-1}.
\end{eqnarray}

In our procedure $COREGA$, considering the orion-like shell structure, the path of photons can be understood in the schematic diagram \ref{fig:rad}. Different layers have different densities and temperature, then different optical depth $\tau_{\nu}$. Hence, the radiation is like a iterative process shown in the following equation:
\begin{eqnarray}
\label{eq:irte}
I_{\nu}^{i}=I_{\nu}^{i-1}e^{-\tau_{\nu}^{i}}+B_{\nu}^{i}(T^{i})(1-e^{-\tau_{\nu}^{i}}).
\end{eqnarray}
where $i$ stands for the $i_{th}$ layer the photons have passed. For the first layer passed through, the specific intensity $I_{\nu}^{0}$ is given by,
\begin{eqnarray}
I_{\nu}^{0}=B_{\nu}^{0}(T^{0})(1-e^{-\tau_{\nu}^{0}}).
\end{eqnarray}

The optical depth for grains of a given type is shown by the equation (\ref{eq:tau}), which is also used in our procedure:
\begin{eqnarray}
\label{eq:tau}
\tau(\nu)=n_d\pi r^2Q(\nu)L,
\end{eqnarray}
where $n_d$ is the number density of dust grains per unit volume in the core, $r$ is the grain radius, $Q$ is the extinction efficiency, and $L $ stands for the path length. $n_d$ is connected with the molecular hydrogen number density $n_{H_{2}}$, through mass ratio of gas to dust $g$. The molecular hydrogen density is one profile we try to fit in the procedure.
\begin{eqnarray}
\label{eq:q}
Q(\nu)=Q_{350}(\frac{\lambda}{350})^\beta,
\end{eqnarray} 
where $\lambda$ is in the unit of $\mu m$, and $Q_{350}$ is the absorption efficiency at $350\mu m$, with the value of $1.36\times 10^{-4}$. $\beta$ is the emissivity spectral index, one important parameter we fit in the procedure.


\end{document}